\documentclass[twocolumn,prb,floats,showpacs,superscriptaddress]{revtex4}
\usepackage{graphicx,epsfig}
\usepackage{times}
\usepackage{graphics,dcolumn,bm,float}
\usepackage{amssymb,amsmath,rotate,color}

\begin{document}
\unitlength 1 cm
\newcommand{\be}{\begin{equation}}
\newcommand{\ee}{\end{equation}}
\newcommand{\bearr}{\begin{eqnarray}}
\newcommand{\eearr}{\end{eqnarray}}
\newcommand{\nn}{\nonumber}
\newcommand{\vk}{\vec k}
\newcommand{\vp}{\vec p}
\newcommand{\vq}{\vec q}
\newcommand{\vkp}{\vec {k'}}
\newcommand{\vpp}{\vec {p'}}
\newcommand{\vqp}{\vec {q'}}
\newcommand{\bk}{{\bf k}}
\newcommand{\bp}{{\bf p}}
\newcommand{\bq}{{\bf q}}
\newcommand{\br}{{\bf r}}
\newcommand{\bR}{{\bf R}}
\newcommand{\up}{\uparrow}
\newcommand{\down}{\downarrow}
\newcommand{\fns}{\footnotesize}
\newcommand{\ns}{\normalsize}
\newcommand{\cdag}{c^{\dagger}}

\title{Enhancement of Rashba spin-orbit coupling by electron-electron interaction}
\author{Rouhollah Farghadan}\email{rfarghadan@kashanu.ac.ir}
\affiliation{Department of Physics, University of Kashan, Kashan,87317-51167, Iran }
\author{Ali Sehat}
\affiliation{Department of Physics, University of Kashan, Kashan,87317-51167, Iran }
\date{\today}

\begin{abstract}
We studied how the electron-electron ($e$-$e$) interaction enhances the strength of the Rashba spin-orbit coupling $(RSOC)$ and opens the possibility of generating a spin-polarized output current from an unpolarized electric current without any magnetic elements. In this regard, we proposed a novel design of a graphene-like junction based on the $e$-$e$ interaction and the $RSOC$. The results showed that the $e$-$e$ interaction with $RSOC$ opens a spin energy gap and extremely enhances the $RSOC$ in this spin energy gap. Interestingly, the junction produces a large spin-polarized current and could act as a high-efficiency spin filter device (nearly $100\%$ spin polarization) at room temperature even at low Rashba and $e$-$e$ interaction strengths. However, with an appropriate design, we showed that $RSOC$ could strongly weaken the spin-polarized current and the spin energy gap, which is solely produced by the $e$-$e$ interaction, but a high spin polarization persists in some energy ranges.
\end{abstract}

\maketitle
\section{Introduction}
Generation of spin-polarized currents and pure spin currents are two important topics in spintronics \cite{Meir,Farghadan1}. All-electrical control of the spin-polarized currents is a basic principle of the operation of novel spintronic devices including the spin field-effect transistors\cite{Manchon,Chuang,Cao,Bhandari}. Historically, Datta and Das \cite{Datta} proposed the first design for applications of the electric field in spintronics \cite{Chappert,Awschalom}. According to this idea the current of spin-polarized electrons that are injected from the ferromagnetic source contact into a semiconductor is modulated by the Rashba spin-orbit coupling ($RSOC$) in a conduction channel\cite{Bychkov,Rashba}.

The problem lies in the ferromagnetic materials used as the source and drain electrodes. These electrodes invariably induce magnetic fields in the channel, seriously impacting spin transport \cite{Cahay1,Cahay2}. Furthermore, the ferromagnetic-semiconductor interface makes spin injection (and extraction) very inefficient. This is the so-called conductivity mismatch problem \cite{Schmidt}.
Therefore, in order to overcome the conductivity mismatch problem in spin transistor devices, the quantum point contacts (QPCs) have been proposed as spin injectors/detectors \cite{Chuang,Thomas,Bauer,Iqbal,Debray,Quay,Rycerz,Tsai}. QPC in graphene \cite{Rycerz} and silicene \cite{Tsai} can be used as a tunable spin polarizer with a nearly perfect spin polarization. The lateral spin-orbit interaction with QPC was also used to control electron spins in a purely electrical manner\cite{Chuang}.

Graphene is also a promising candidate for spintronics applications
owing to its small intrinsic spin-orbit and high electron mobility
\cite{Hernando,Neto}. In graphene, there is an intrinsic spin-orbit coupling that was
first studied by Kane and Mele\cite{Kane2,Kane1}. Moreover, the extrinsic
$RSOC$, which is originated from the inversion asymmetry that can be caused
by a perpendicular electric field, the interaction with substrate
and the atoms adsorbed on the surface, was studied
\cite{Kane1,Varykhalov,Neto2,Abdelouahed,Min}. There are many recent
theoretical \cite{Kane2,Kane1,Zarea,Liu2} and experimental
\cite{Varykhalov,Dedkov} works on the role of the Rashba term in the
electronic structure of graphene. Recent experiments have shown that
the strength of $RSOC$ for graphene can reach values up to 200 meV
\cite{Dedkov}. In addition, other plan honeycomb structures such as
silicene and germanene are good candidates for spintronic
applications due to their long spin-relaxation times and lengths
and edge magnetism \cite{Cahangirov}.

The $e$-$e$ interaction has a significant effect on electronic, magnetic and transport properties of these structures. The zigzag edges of these graphene-like structures are locally magnetized due to the $e$-$e$ interaction\cite{Farghadan2,Magda,Cahangirov,Wang}.
The simultaneous effects of $RSOC$ and $e$-$e$ interaction are previously studied\cite{Wan,Pournaghavi,Sun,Lopez}. Previous works have showed that \cite{Sun} the spin-orbit and $e$-$e$ interactions could strengthen the spin accumulation. Also, the transport properties of a quantum wire with a local Rashba interaction in the presence of a coulomb repulsion predict Kondo regime\cite{Lopez}.

In this paper, by using $e$-$e$ interaction we proposed a design to
efficiently enhance the $RSOC$ in graphene-like junction at room
temperature. We proposed all-electrical spin filter devices without any
ferromagnetic electrodes and any magnetic fields, which could
overcome the conductivity mismatch problems and their associated inefficiencies. The $e$-$e$
interaction in zigzag edge nanoribbons served as electrodes in the
presence of $RSOC$ in a small part of the junction served as the channel
could generate a large spin-polarized current with a high degree of spin
polarization at room temperature. In fact, $e$-$e$ interaction with
$RSOC$ opens a spin energy gap and enhances the $RSOC$ in the
junction.  We examined the spin-dependent transport features of these
devices for various $e$-$e$ and Rashba strengths and found a nearly
perfect spin filtering effect even at low Rashba and $e$-$e$ interaction strengths . The results
showed that with an appropriate design, the $RSOC$ could strongly
weaken the spin-polarized current and the spin energy gap, which is produced by the $e$-$e$ interaction. This result may be utilized for two dimensional (2D) honeycomb
structures such as graphene, silicene and germanene.

\section{MODEL AND METHOD}

We consider a system that consists of an armchair-edge graphene flake
with a finite length, sandwiched between two perfect semi-infinite
zigzag graphene nanoribbons as electrodes. To calculate the
spin-dependent transport properties of the graphene flake, which acts as
the channel, we decompose the total Hamiltonian of the system as:
\begin{equation}
 \hat{H} =\hat{H_L}+\hat{H_C}+\hat{H_R}+\hat{H_T}
\end{equation}
where $\hat{H_L}$ and $\hat{H_R}$ are the Hamiltonians of the left (L) and right (R) electrodes, respectively, where $\hat{H_C}$ describes the channel Hamiltonian and contains the $RSOC$ term, and $\hat{H_T}=-t\sum_{i,\alpha} c^{\dagger}_{i,\alpha}d_{j,\alpha}$ is the coupling between the electrodes and the central part (channel); and t is the tight binding parameter which will be
set to t = 2.66 $eV$.

The Hamiltonian of zigzag graphene nanoribbons ($\hat{H_E}=L,R$), which contains the $e$-$e$ interaction, within the single single-band tight-binding approximation and the mean field Hubbard model for a finite temperature can be written as:
\begin{equation}
\hat{H_E}=\sum_{<i,j>,\sigma}{[(\epsilon_{i,\sigma}-\mu)\delta_{ij}-t]~ d^{\dagger}_{i,\sigma}d_{j,\sigma}}
+U\sum_{i,\sigma}\hat{n}_{i,\sigma}\langle \hat{n}_{i,-\sigma}\rangle \,
\end{equation}
where $\epsilon_{i,\sigma}$ is the on-site energy, $\mu$ is the chemical potential and $t$ is the tight-binding parameter which will be set to $t=2.66$ $eV$. In this expression, $d^{\dagger}_{i\sigma}$ ($d_{i\sigma}$) creates (annihilates) an electron, and $\langle \hat{n}_{i,\sigma}\rangle$ is the electron density of the number operator for an electron with spin $\sigma$ at the $ith$ site. The latest term accounts for the on-site Coulomb interaction $U$. The electron densities and chemical potential are calculated self-consistently by the iteration method\cite{Magda,Farghadan2}. Conservation of the number of electrons defines the chemical potential at a finite temperature. Therefore, the chemical potential and the electron densities for zigzag nanoribbon should be calculated iteratively until a convergence of the electron density is reached. The iteration is stopped when the difference between two successive iterations becomes less than $10^{-6}$. Moreover, the magnetic moment at each site can be expressed as: $m_i=\mu_B(\langle \hat{n}_{i,\sigma}\rangle-\langle \hat{n}_{i,-\sigma}\rangle)$.

The Hamiltonian of the central part with $RSOC$ term and without $e$-$e$ interaction can be written as \cite{Chico}:
\begin{equation}
\hat{H_C}=-t\sum_{i,\alpha} c^{\dagger}_{i,\alpha}c_{j,\alpha}
+i\frac{\lambda}{a}\sum_{<i,j>,\alpha,\beta}c^{\dagger}_{i,\alpha}{(\sigma\times d_{ij})}_{z,\alpha\beta}c_{j,\beta}
\end{equation}
The first term is the single-band tight-binding Hamiltonian, and the second term stands for the $RSOC$. $\lambda$ is the Rashba strength, $d_{ij}$ is the displacement vector between sites $i$ and $j$, and the indexes of $\alpha,\beta=\uparrow,\downarrow $, $a$ is carbon-carbon bond length and $\sigma$ is the Pauli matrix.

According to the Landauer-Buttiker formalism the spin-dependent conductance can be written as \cite{Farghadan3}:
\begin{equation}
G^{LR}_{\alpha\beta}(\varepsilon)= \frac{e^2}{h} {Tr}[\hat{\Gamma}_{\alpha}(\varepsilon)
\hat{G}_{\alpha\beta}(\varepsilon)\hat{\Gamma}_{\beta}(\varepsilon)\hat{G}_{\beta\alpha}^{\dagger}(\varepsilon)].
\end{equation}
Where $\hat{G}_{\alpha\beta}$ is the Green's function of the junction. Using the self-energy matrices due to the connection of electrodes to the channel $\hat\Sigma_{\alpha}(\varepsilon)$, the coupling matrices $\hat{\Gamma}_{\alpha}(\varepsilon)$ can  be expressed as $\hat{\Gamma}_{\alpha}(\varepsilon)=-2 Im[\hat\Sigma_{\alpha}(\varepsilon)]$.
Moreover, $ {G}_{\alpha\beta}(\varepsilon)$ is the probability that an electron from the left electrode with the $\alpha$-spin direction and $(\varepsilon)$ energy is detected in the right electrode with the $\beta$ spin direction. Moreover, $P_z(\varepsilon)$ the spin polarization in $z$ direction can be defined as:

\begin{equation}
P_z(\varepsilon)=\frac{{G}_{\uparrow\uparrow}(\varepsilon)+{G}_{\downarrow\uparrow}(\varepsilon)-{G}_{\downarrow\downarrow}(\varepsilon)-{G}_{\uparrow\downarrow}(\varepsilon)}{{G}_{\uparrow\uparrow}(\varepsilon)+{G}_{\downarrow\uparrow}(\varepsilon)+{G}_{\downarrow\downarrow}(\varepsilon)+{G}_{\uparrow\downarrow}(\varepsilon)}\times 100
\end{equation}

\section{RESULTS AND DISCUSSION }

\begin{figure}
\centerline{\includegraphics[width=1\linewidth]{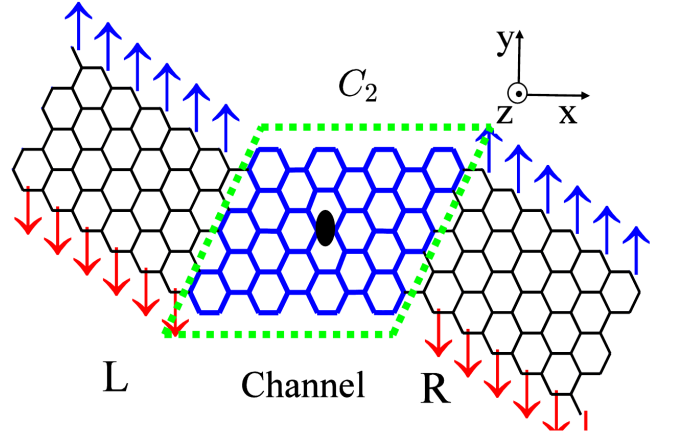}}
\caption{(a) Schematic view of parallelogram-shaped junction with localized magnetic moments on zigzag edges. The $\uparrow$ ($\downarrow$) correspond to the majority (minority) spin electrons. Two zigzag graphene nanoribbons as left (L) and right (R) contacts which connected to parallelogram-shaped channel (green dashed region with $RSOC$). $C_2$  two-fold rotational symmetry, also the magnetic moments in contacts are within [-0.15:0.15].
}
\end{figure}

We investigate the spin-dependent transfer features of 2D honeycomb structures by mean field Hubbard model and tight-binding approximation in the presence of Rashba spin orbit interaction. Fig. 1 shows that a parallelogram-shaped graphene channel with its armchair edges is sandwiched between two perfect semi-infinite zigzag graphene nanoribbons (electrodes). The armchair parallelogram-shaped channel consists of 72 carbon atoms and does not show any magnetic moments at the edges. Therefore, we do not consider the $e$-$e$ interaction in the channel. Also, the Rashba spin-orbit interaction only occurs in the channel region, which is produced by an electric field appiled perpendicularly on the honeycomb plane ($z$ direction). By choosing an appropriate design for the junction, the left and right semi-infinite zigzag nanoribbons have parallel spin configurations. Moreover, each electrode has an antiferromagnetic spin configuration and is ferromagnetically coupled to the other one (see Fig. 1). Moreover, the $e$-$e$ interaction is only considered in left and right contacts due to zigzag edge states.

This graphene junction is not similar to the spin-dependent Datta-Das
transistor. Unlike the Datta-Das spin transistor, we use zigzag
nanoribbons as electrodes, which could not inject the spin-polarized
current in the channel. In fact, we remove the ferromagnetic contacts to
overcome the conductivity mismatch problem. The channel has a two-fold rotational
symmetry around $z$ direction ($C_2$). Due to this symmetry the
spin-conserved conductances
(${G}_{\uparrow\uparrow}={G}_{\downarrow\downarrow}$) are the same
and only the spin-flipped conductances
(${G}_{\uparrow\downarrow}\neq{G}_{\downarrow\uparrow}$) are
different from each other \cite{Chico}. Therefore, spin polarization
can be induced in this junction with the spin-flipped conductance
alone.

We first study the transport properties of a parallelogram-shaped junction without $e$-$e$ interaction. In Fig 2(a), we plotted the spin-flipped conductance with $\lambda=0.2$ $eV$ and without $e$-$e$. There is a little spin splitting between the two spin-flipped conductances, leading to a finite spin polarization as shown in Fig 2(c).
Moreover, due to the antiferromagnetic spin configuration of zigzag nanoribbons, the $e$-$e$ interaction alone cannot produces a spin-polarized current in the absence of  $RSOC$ in the junction.

\begin{figure}
\centerline{\includegraphics[width=.95\linewidth]{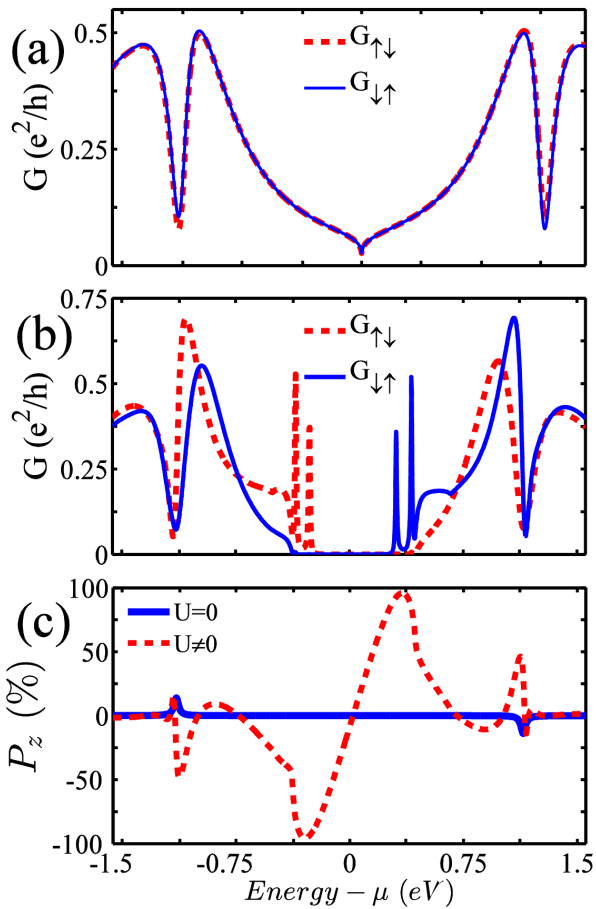}}
\caption{ Spin-resolved conductance and spin polarization of the parallelogram-shaped junction. (a) Spin-flipped conductance without considering $e$-$e$ interaction in contacts ($U=0$), and with $\lambda=0.2$ $eV$. (b) Spin-flipped conductance for $U=2.82$ $eV$ and $\lambda=0.2$ $eV$. (c) Shows spin polarization as a function of the energy, the dashed and solid line for $U=0$ and $U=2.82$ $eV$, respectively.}
\end{figure}

In order to understand the effect of $e$-$e$ interaction with $RSOC$
on the generation of spin-polarized current, we plotted the
spin-flipped conductance as a function of energy for $U=2.82$ $eV$
and $\lambda=0.2$ $eV$ at room temperature, as shown in Fig 2(b).
Interestingly, a large spin-polarized current appears in the energy
ranges from $-.35$ $eV$ to $-.25$ $eV$ and from $.25$ $eV$ to $.35$
$eV$ (see Fig. 2(b)), which is the spin energy gap induced by $e$-$e$
and Rashba spin-orbit interactions. Moreover, the spin energy gap is
opened and the degeneracy between the two spin-flipped conductances is
removed. Therefore, the channel produces a large spin-polarized
current with a high degree of spin polarization in a large energy
region, which is depicted in Figs. 2 (b) and 2(c). Note that for two
cases ($U=0$, $U\neq0$) the charge current through the zigzag-edge
electrodes is not spin-polarized, but when the electron arrives at
the Rashba region (a region where the $RSOC$ is present), the Rashba field causes the spin of the injected electron precesses in the channel.

Moreover, the $e$-$e$ interaction greatly enhances the precession of the electron spin and makes the junction a perfect spin filter device. Generally, $e$-$e$ interaction with $RSOC$ in a special design (parallelogram-shaped junction) opens a spin energy gap, generating a large spin-polarized current in the spin energy gap at room temperature. Similarly, the competition between Zeeman effect and the spin-orbit interaction and the interplay between Rashba spin-orbit interaction and local magnetic moments can open a spin energy gap with a fully spin-polarized current in the spin energy gap\cite{Streda,Zhai}. Moreover, a spin energy gap in the ferromagnetic graphene junction and the magnetic barrier in the presence of strain or $RSOC$ can produce a fully spin polarized current\cite{Wu,Wu2}.

\begin{figure}
\centerline{\includegraphics[width=.96\linewidth]{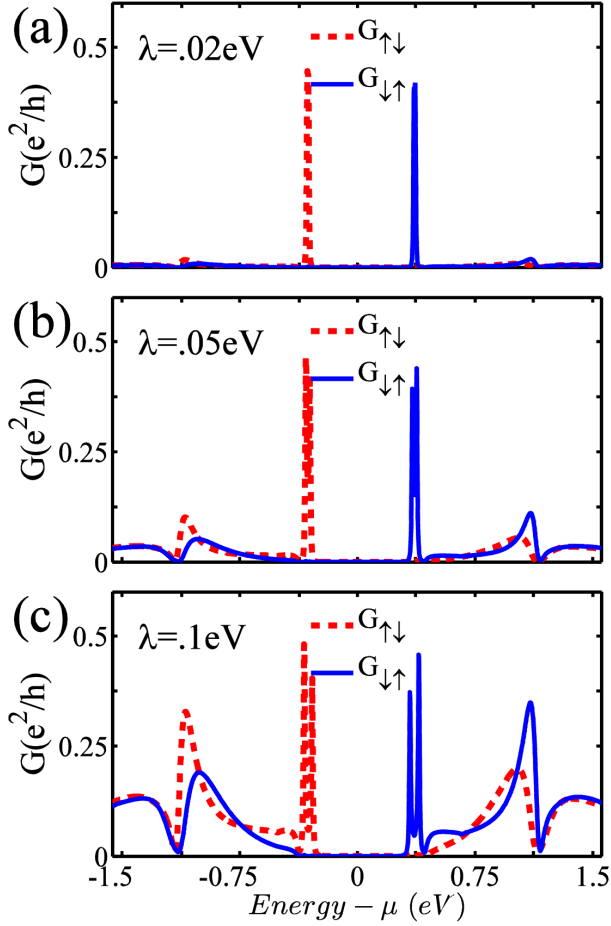}}
\caption{ Spin-resolved conductance of the parallelogram-shaped junction  for various Rashba strengths and with $U=2.82$ $eV$.}
\end{figure}

\begin{figure}
\centerline{\includegraphics[width=.96\linewidth]{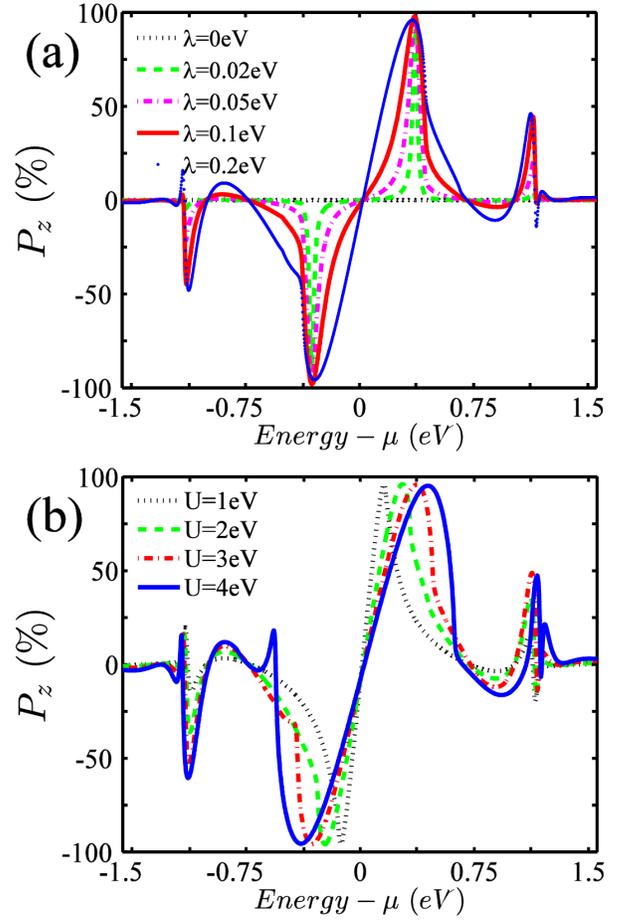}}
\caption{ Spin polarization as a function of the energy for various Rashba and $e$-$e$ interaction strengths. (a) for $U=2.82$ $eV$ and (b) for $\lambda=0.2$ $eV$}
\end{figure}

In order to see the sensitivity of the graphene junction to the $RSOC$ strength, we plotted the spin-flipped conductance as a function of energy for various Rashba strengths, as shown in Fig. 3. Comparing the spin-resolved conductances for different strengths of $RSOC$ shows that as the Rashba strength increases the peaks of the spin-flipped conductance , ${G}_{\uparrow\downarrow}$ and ${G}_{\downarrow\uparrow}$, near the Fermi energy ($E$=-0.3 $eV$ $E$ = 0.3 $eV$ )  become wider, but the magnitude of these peaks remain unchanged. Moreover in some energy ranges far from Fermi energy, about 1$eV$ and -1$eV$, as the Rashba strength increases the magnitude of spin-flipped conductances enhance. Generally, in the presence of $e$-$e$ interaction as the Rashba strength increases the spin energy gap and the spin precession in the parallelogram-shaped junctions increase (see Fig. 3). These results may be useful for the design of spintronic devices based on two dimensional honeycomb structure.

We also plotted the polarization component $P_z$ as a function of energy for various Rashba strengths, as shown in Fig. 4(a). Interestingly a nearly full spin polarization can be generated for all $RSOC$ strengths in the spin energy gap. Moreover, as the Rashba strength increases, the position of the maximum spin polarization remains unchanged. However, as the $RSOC$ strength increases the $P_z$ function becomes wider in the spin gap energy. Generally, as the Rashba strength increases the spin-polarized current and the spin energy gap in the parallelogram-shaped junctions increase.

In order to study the effect of $e$-$e$ interaction strength on transfer features of the graphene junction, the spin polarization for various strengths of $e$-$e$ interaction is plotted in Fig. 4(b). The results show that as  the $e$-$e$ interaction strength increases, the spin energy gap width and spin-polarized current (Not shown here) increase. Moreover, for all values of $e$-$e$ interaction strength, a nearly full spin polarization produces, but the position of the maximum spin polarization varies. Therefore, as the $e$-$e$ interaction strength increases, a larger spin energy gap width is produces with a larger spin-polarized current. Note that because of the electron-hole symmetry property of  conductance, the conductance as a function of energy satisfies ${G}_{\uparrow\downarrow}(-\varepsilon)={G}_{\downarrow\uparrow}(\varepsilon)$, and so $P_z(-\varepsilon)=-P_z(\varepsilon)$.

Clearly, $e$-$e$ interaction produces edge magnetism and energy gap in zigzag graphene nanoribbons. We showed that due to $e$-$e$ interaction with $RSOC$, this energy gap opens a spin energy gap and enhances the Rashba strength in the graphene junction. Recently, Magda et al. have showed that the energy gap and edge magnetism for zigzag ribbons that are narrower than 7 nm in width can be stable by doping and in the presence of edge irregularities\cite{Magda} and edge roughness\cite{Farghadan3} even at room temperature . On the other hand, armchair and zigzag graphene nanoribbons can be fabricated with a precise edge orientation in the nanometer scale based on various nanofabrication techniques such as scanning tunneling microscopy \cite{Magda}. Moreover, graphene based spin devices with $RSOC$ can be manipulated at room temperature \cite{Cahangirov}. Therefore, our theoretical results may be useful for practical engineering of spintronic devices with edge irregularities, roughness and doping for nanoribbon junctions that are narrower than 7 nm in width at room temperature.

\begin{figure}
\centerline{\includegraphics[width=1\linewidth]{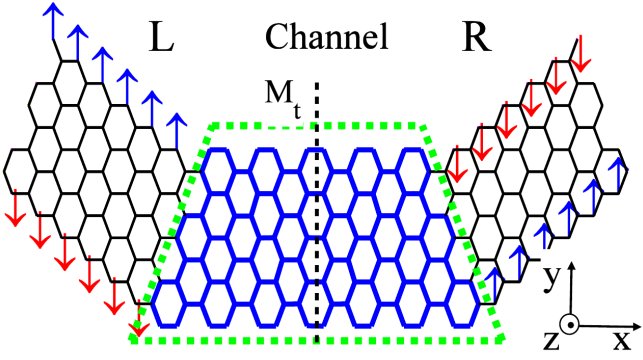}}
\caption{(a) Schematic view of trapezoidal-shaped junction with localized magnetic moments on zigzag edges. The $\uparrow$ ($\downarrow$) correspond to the majority (minority) spin electrons. Two zigzag graphene nanoribbons as left (L) and right (R) contacts which connected to trapezoidal channel (green dashed region with $RSOC$).$M_t$ is transversal mirror symmetry and the magnetic moments in contacts are within [-0.15:0.15].
}
\end{figure}

\begin{figure}
\centerline{\includegraphics[width=0.95\linewidth]{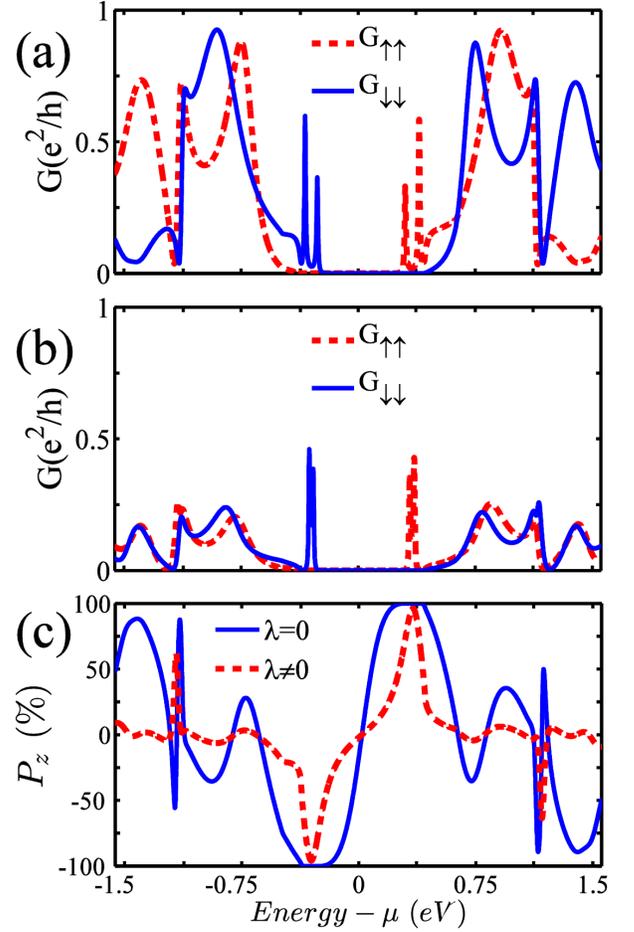}}
\caption{ Spin-resolved conductance and spin polarization of the trapezoidal-shaped junction. (a) Spin-conserved conductance for ($U=2.82$ $eV$) and for $\lambda=0 $. (b) Spin-conserved conductance for $U=2.82$ $eV$ and $\lambda=0.2$ $eV$. (c) Shows spin polarization as a function of the energy for $ U=2.82$ $eV$.
}
\end{figure}

In order to investigate the mutual effect of $RSOC$ and $e$-$e$ interaction, we proposed the trapezoidal-shaped graphene junction that is depicted in Fig. 5. Under $e$-$e$ interaction, this junction produces a large spin-polarized current with a high spin polarization without Rashba spin-orbit interaction. We investigated the effect of Rashba field on spin polarization induced by $e$-$e$ interaction. In Fig. 5, a trapezoidal-shaped graphene channel with armchair edges is sandwiched between two perfect semi-infinite zigzag graphene nanoribbons (electrodes). The armchair trapezoidal-shaped channel consists of 108 carbon atoms. By choosing an appropriate design for the junction, the left and right semi-infinite zigzag nanoribbons have antiparallel spin configurations. The zigzag honeycomb ribbons have two different sublattices at the outermost atoms. We define the outermost atom on the top of the left electrode ($L$) as sublattice $A$ and the outermost atom on the bottom as sublattice $B$. However, the sublattice on the right electrode ($R$) is reversed. Therefore, the magnetic edge states in both electrodes are coupled in the antiparallel manner, rather than  parallel. Moreover, each electrode has an anti-ferromagnetic spin configuration and is antiferromagnetically coupled to the other one, (see Fig. 5).

We examined the spin-dependent transport features of the junction
with $e$-$e$ interaction and without $RSOC$, (see Fig. 6(a)). This
mismatching between the two sublattices in the left and right electrodes
opens a spin energy gap from $-.55$ $eV$ to $-.25$ $eV$ and from
$.25$ $eV$ to $.55$ $eV$ and causes a large spin-polarized current in
the spin energy gap. The spin-conserved conductance shows that the junction
acts as a perfect spin filter device. Note that, the trapezoidal-shaped
graphene channel has the transversal mirror symmetry $M_t$. Due to this
symmetry in the presence of $RSOC$, spin-flipped conductances
(${G}_{\uparrow\downarrow}={G}_{\downarrow\uparrow}$) are the same
and only the spin-conserved conductances
(${G}_{\uparrow\uparrow}\neq{G}_{\downarrow\downarrow}$) are
different from each other\cite{Chico}. Therefore, a spin polarization
can be induced in the trapezoidal-shaped junction with the spin-conserved
conductances alone.

Next, we investigated the effect of $RSOC$ on the spin-resolved
conductance, which is solely induced by $e$-$e$ interaction. We plotted the
spin-conserved conductances for the trapezoidal-shaped junction with the $RSOC$
and $e$-$e$ interaction (see Fig. 6(b)). Clearly, $RSOC$ in
the trapezoidal-shaped junction extremely reduces the spin-polarized current which
is produced by the $e$-$e$ interaction. The $RSOC$ causes precession of
electron spin. Moreover, in the trapezoidal-shaped graphene junction the
precession of electron spin extremely reduces the height and
width of the function of spin-conserved conductances. Therefore, the
decrease in the conductance due to the $RSOC$  leads to a decrease in the spin energy gap (see
Fig. 6(b)). But a high spin polarization persists in some energy ranges,
even with the precession effect of $RSOC$ on the spin-polarized current
of $e$-$e$ interaction (see
Fig. 6(b)). Moreover, this decrease depends on the
$RSOC$ strength and a lower Rashba strength induces a lower decrease
in the spin-conserved conductance. Generally, $RSOC$ could
strongly weaken the spin-polarized current and the spin energy gap, which is produced by $e$-$e$
interaction. But a high spin polarization persists in some energy
ranges (see Fig. 6(c)).

Finally, our results can be applied for another 2D honeycomb structures such as silicene and germanene. Silicene and germanene are two important 2D graphene-like systems. The honeycomb structure of these materials consists of two sublattices. Armchair and zigzag nanoribbons of silicene and germanene are stable and have electronic and magnetic properties similar to graphene. The ground state of zigzag nanoribbons has an antiferromagnetic spin configuration with an energy gap \cite{Cahangirov, Wang}. Also, these structures, due to their buckled structures, have larger spin-orbit interactions than those of graphene \cite{Liu}. Interestingly, silicene has a longer diffusion length and time  than those of graphene and is more appropriate for current Si-based devices\cite{Wang}. Therefore, we predict that silicene and germanene structures similar to graphene show high spin polarization effects due to the enhancement of the $RSOC$ by $e$-$e$ interaction.

\section{Conclusion}
We investigated how the $e$-$e$ interaction enhances the $RSOC$ in a honeycomb two dimensional junction. Based on $e$-$e$ interaction and $RSOC$, we proposed an all-electrical novel spin filter device to overcome the conductivity mismatch problem. The results showed that $e$-$e$ interaction and Rashba strengths opens a spin energy gap and can induce a high spin polarization in the junction at room temperature. Therefore, the junction acts as a perfect spin-filter device with a large spin-polarized current in the spin energy gap. Generally, a high spin polarization can be found even the Rashba and $e$-$e$ interaction strengths are low at room temperature. However, $RSOC$ could strongly weaken the spin-polarized current, which is solely produced by $e$-$e$ interaction, but a high spin polarization persists in some energy ranges. Generally, in parallelogram-shaped junctions, as the Rashba and $e$-$e$ interaction strengths increase the spin-polarized current and spin energy gap increase too. Moreover, our theoretical results may be useful for practical engineering of spintronic devices at ambient conditions based on two-dimensional honeycomb structures such as graphene, silicene, and germanene.

\end{document}